# Multipole engineering for enhanced backscattering modulation

Dmitry Dobrykh[1,2,a),b)], Diana Shakirova[1,a)], Sergey Krasikov[1], Anna Mikhailovskaya[1,2], Ildar Yusupov[1], Alexey Slobozhanyuk[1], Konstantin Ladutenko[1], Dmitry Filonov[3], Andrey Bogdanov[1] and Pavel Ginzburg[2]

[1)] *Department of Physics and Engineering, ITMO University, Saint Petersburg 197101, Russia*
[2)] *School of Electrical Engineering, Tel Aviv University, Tel Aviv 69978, Israel*
[3)] *Center for Photonics and 2D Materials, Moscow Institute of Physics and Technology, Dolgoprudny 141700, Russia*
a) D. Dobrykh and D. Shakirova contributed equally to this work
b) d.dobrykh@metalab.ifmo.ru

*Abstract* - An efficient modulation of backscattered energy is one of the key requirements for enabling efficient wireless communication channels. Typical architectures, being based on either electronically or mechanically modulated reflectors, cannot be downscaled to subwavelengths dimensions by design. Here we show that integrating high-index dielectric materials with tunable subwavelength resonators allows achieving an efficient backscattering modulation, keeping a footprint of an entire structure small. An interference between high-order Mie resonances leads to either enhancement or suppression of the backscattering, depending on a control parameter. In particular, a ceramic core-shell, driven by an electronically tunable split ring resonator was shown to provide a backscattering modulation depth as high as tens of the geometrical cross-section of the structure. The design was optimized towards maximizing reading range of radio frequency identification (RFID) tags and shown to outperform existing commercial solutions by orders of magnitude in terms of the modulation efficiency. The proposed concept of multipole engineering allows one to design a new generation of miniature beacons and modulators for wireless communication needs and other relevant applications.

Modulation of electromagnetic signals is a cornerstone for enabling a vast majority of nowadays communication systems. Information is typically encoded by imposing amplitude and phase temporal changes on a carrier wave. Specific application requirements dictate critical parameters, such as the carrier frequency and the bandwidth, essential to ensure a sufficient level of transmitted energy and information capacity of a channel. Carrier frequencies can be as small as several Hz (e.g. for submarine communications[1,2,] and grow up to hundreds of THz in optical communications[3] and even further towards X-rays to serve frontier space missions.[4] While numerous modulator architectures have been developed to serve the above-mentioned needs, they all share the same property – relatively bulky footprint with respect to the central operational wavelength.[5] The physical reason here relies on the essence of electromagnetic field interactions with matter. To impose a significant phase shift on a propagating wave, an interaction length should prevail at a distance over a wavelength. For example, typical telecommunication optoelectronic modulators, based on Mach-Zehnder interferometer architecture, are as big as several millimeters.[6]

To achieve a significant interaction between waves and matter in a small volume, resonant phenomena should be employed. Theoretically, a $\pi$-shift can be achieved via an interaction with a high-quality resonator by tuning its resonant frequency around the carrier one. This single resonance approach has an immediate drawback, tightly related to the well-known Chu-Harrington limit,[7] which predicts a significant drop in an operational bandwidth with a device footprint reduction. An overview of the beforehand mentioned constrains indicates that obtaining a significant phase and amplitude modulation within a small interaction volume without a significant bandwidth degradation is a challenging task, which requires the development of new theoretical and experimental approaches. A possible solution to the problem is to employ multiple resonances in a small structure, as we will show hereinafter.

To obtain a resonant response in a subwavelengths structure, the later should be made of a material with high permittivity (we will put localized plasmon resonance phenomena aside).[8,9] The concept of dielectric resonant antennas (DRAs) has been developed in the radio physics community and was found to be beneficial in cases, where e.g. size reduction without a significant bandwidth degradation or sustainability to high radiated power are needed.[10,11] Typically, high-quality ceramics are used in those applications.[12,13] Ceramic composites can have electric permittivities as high as hundreds and thousands without exhibiting significant ohmic losses. Recently, the concept of resonance engineering in high-index dielectric materials has been pursued. For example, all-dielectric nanoantennas and metasurfaces have contributed to the tailoring of light-matter interaction on a nanoscale.[14] A vast majority of designs are based on controlling electric and magnetic Mie resonances in spheres, cylinders, and other shapes, which can be accurately fabricated from semiconductor wafers using different lithographic techniques or workshop post-processing in the case if cm-scale devices.[15–18]

While quite a few different applications of all-dielectric electromagnetism (i.e. DRAs and nanophotonics) have been discussed,[13,19] here we will concentrate on a relatively new aspect, which has both fundamental and applied sides. In particular, the radiation cross-section (RCS) and back-scattered modulation efficiency will be discussed. It is worth noting that this parameter is a key in applications, where beacon-like devices are in use. The well-known examples include friend or foe transponders, marine radar reflectors, radio frequency identification (RFID), and many others. The design of beacons with a small footprint is the challenge, which will be addressed here. To make the report less abstract and more applied, we will focus on RFID applications.

RFID is a widely used form of wireless communication that is based on time-modulated electromagnetic or electrostatic coupling to uniquely identify an object.[20] A typical RFID system consists of passive tags and an active reader, which trigger tags using interrogation electromagnetic pulse. The readout of digital data is performed by analyzing time-modulated backscattered signals. High-frequency RFID systems operate at 860-960 MHz frequency bands, making free-space propagation mechanisms to govern the physics of the communication channel. This far-field approximation is even better justified in cases when long-range readout is required. Typical distances, considered as a long-range, are several meters and can reach 20 meters and higher if state-of-the-art equipment and antenna design are employed.[21,22] The key parameters, capable to extend the readout range of a tag are a load factor and modulation efficiency.[23] The first one is related to an energy harvesting to tag's electronics, while the second one is capable of establishing an efficient communication channel – the beforehand discussed beacon application. In a monostatic regime, where a single interrogating antenna is used, the power modulation of the backscattered signal is given by a radar equation as follows:[23]

$$\Delta P^{RCS} = \frac{P_r \lambda^2 G_r^2 \Delta\sigma_{back}}{(4\pi)^3 r^4}, \qquad (1)$$

where $P_r$ is the power transmitted by a reader, $G_r$ is the gain of the reader's antenna, $r$ is the distance from the reader to the tag, $\lambda$ is the carrier's wavelength, and $\Delta\sigma_{back}$ is a modulated backward cross-section of a tag. Since the value of $(P_r + G_r)$ in dBm cannot exceed 36 dBm (4 W of Equivalent Isotropically Radiated Power) according to international regulations, $\Delta\sigma_{back}$ becomes the key parameter for optimization. Typically, RFID communication protocols are based on amplitude-shift keying (ASK) and, hence, both minimal and maximal values of backscattering are important to maintain a robust continuous communication.[24]

Furthermore, an additional valuable property is anisotropy of a readout scenario. Standard RFID systems, based on dipole-like antenna tags, suffer from a polarization mismatch – a misalignment of tag's orientation with respect to the reader's antenna, which results in the readout distance drop. Creating isotropic beacons is an additional important task, which allows addressing this issue. Having in mind those general requirements, we will develop a new approach, based on multimode interference within small high-index resonators.

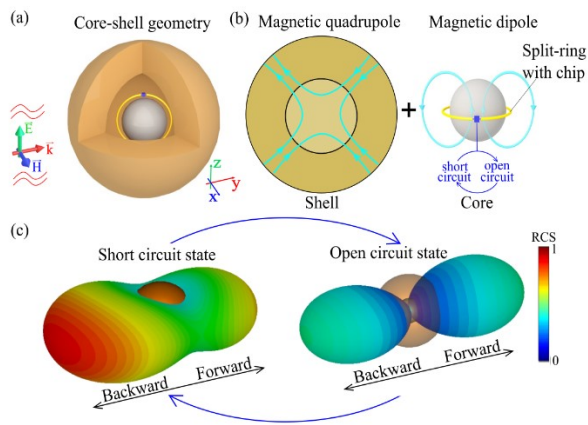

FIG. 1. (a) A core-shell structure for a maximized modulated backward cross-section. (b) The principle of operation – an interference between magnetic quadrupole resonance is in a shell and magnetic dipole is in a core. The dipolar resonance is controlled by an impedance of a split ring resonator (e.g. short/opened circuit or RFID integrated circuit), wrapped around the sphere. (c) Modulation of radiation pattern for two states of the integrated circuit impedance.

To maximize the modulated backward cross-section, we will explore interference between multipoles (Mie resonances),[25] providing the efficient directive scattering. The most pronounced example of scattering manipulation via multipole interference is the Kerker effect, which based on an interplay between electric and magnetic dipoles. Initially, it was proposed to suppress the backscattering.[26] Today this effect has been realized in many different geometries and was generalized to include interference between higher-order multipoles.[27] Here, we will explore a core-shell geometry, optimizing it towards providing a maximized modulated backward cross-section.

Figure 1 demonstrates the concept design of the resonator – a spherical ceramic core and a shell, separated by a transparent (e.g. air) layer. Spherical shape allows associating each resonance of the structure (Mie resonances) with a single multipole (dipole, quadrupole, etc). It is worth noting that multipole mixing in structures, which do not obey spherical symmetry, complicates the analysis.[28] Here, magnetic dipole and magnetic quadrupole were designed to resonate at nearly the same frequency [Fig. 1 (b)]. To tune the spectral position of the magnetic dipole resonance and, as a result, to control the backscattering, the core is functionalized with a split-ring resonator (SRR). This element has a chip with variable impedance, switched between open or short circuit states [see Figs. 1 (b) and 1 (c)]. In practical RFID applications, the switch will be replaced with an integrated circuit, which has a vendor-defined internal impedance, subject to a protocol-based modulation scheme.

Relatively high values of permittivity of core allow manipulating the resonances of the core and shell independently as a first approximation. Exact Mie solutions exist for spherically symmetric structures.[29] Fig. 2 shows color maps of the backward cross-section as a function of the system's parameters. The permittivity of the core and the shell are subject to optimization. The operational frequency was chosen to be 900 MHz, consistently with RFID standards (EPCGEN2 UHF RFID band is 860-960 MHz).[24] Other parameters: the core's radius $r_{core} = 5.5$ mm and inner and outer radiuses of the shell $r_{shell}^{inner} = 13.5$ mm and $r_{shell}^{outer} = 39.5$ mm (there is an air layer between core and shell 8 mm wide). Those parameters were chosen to provide magnetic dipole (MD), magnetic quadrupole (MQ), and electric dipole (ED) resonances of the standalone structures (the shell and the core) in the range of achievable parameters. Mixing powders of high-quality ceramics allows obtaining moderately high values of relative permittivities. The ranges of values, considered in the analysis, are $\varepsilon_{shell} = 33 - 49$ and $\varepsilon_{core} = 900 - 930$. The loss tangent of the materials was taken to be $tan\,\delta = 5 \cdot 10^{-4}$, consistently with reported experimental data.[30]

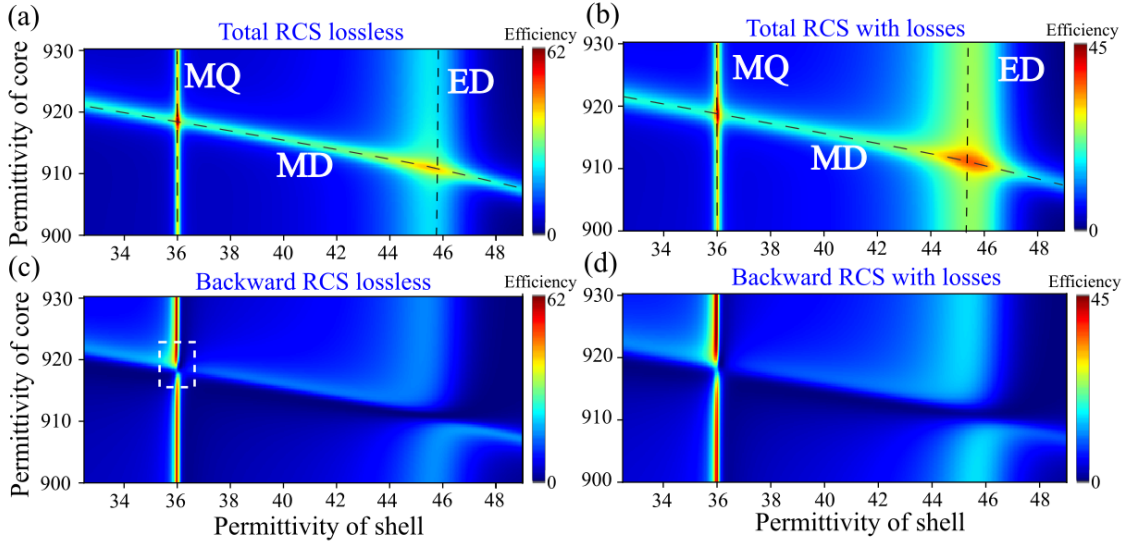

FIG. 2 Scattering cross-section colormaps as the function of cores and shells relative permittivities. Total scattering cross section efficiencies for (a) lossless case, (b) lossy case. Backward scattering cross section efficiencies for (c) lossless and (d) lossy cases. Scattering efficiencies are normalized to the geometrical cross section of the structure. Other system's parameters appear in the main text.

The total and backscattering cross-sections for both lossless and lossy cases appear in Fig. 2 (a-d), where the evolution of resonant branches can be clearly seen. The most interesting points correspond to the intersection of modes, e.g. ED-MD and MQ-MD. The values on the colormaps are normalized to the geometrical cross-section (i.e. $\pi(r_{shell}^{outer})^2$) of the structure and, hence, represent the scattering efficiencies. It is worth noting that the backward scattering efficiency can be as high as 62 (45) in lossless (lossy) cases, which is beneficial for establishing an efficient communication channel with miniature ($\sim\lambda/10$) backscatterers.

In principle, well-known interplaying Kerker and anti-Kerker effects (suppression of backward and minimizing of forward scattering, respectively) can be used for the modulation effect. In this case, the controllable interference between ED and MD resonances can be utilized. However, nulling the backward scattering does not necessarily maximize the modulation efficiency. In particular, minimization of forward scattering, associated with anti-Kerker effect, results in a total scattering cross-section drop, according to the optical theorem.[31]

The most interesting region of parameters for an efficient backscattering modulation is highlighted with a white dashed rectangle in Fig. 2 (c). The destructive interference between MQ and MD leads to the suppression of the backscattering. However, a minor change in the core's permittivity results in a jump of the backscattered signal – this is the work point, which will be analyzed in detail next. It is also worth noting that losses do not cause significant degradation of the effect. $\Delta\sigma_{back}$ reaches values of 41 (26) geometrical cross-section for lossless (lossy) cases.

To highlight the tuning capability, Fig. 2 (d) was sliced along the MQ branch. Shell's permittivity in this case is $\varepsilon_{shell} = 36.61 + 0.01i$. The backscattering efficiency is then plotted as the function of the core's permittivity (Fig. 3). Insets in the figure indicate the far-field scattering diagrams, demonstrating that only 0.5% permittivity tuning leads to $\Delta\sigma_{back}$ of about 26 geometrical cross-sections. It is worth noting that the backscattering does not vanish and remains relatively high, which might be beneficial for ASK communication protocol.

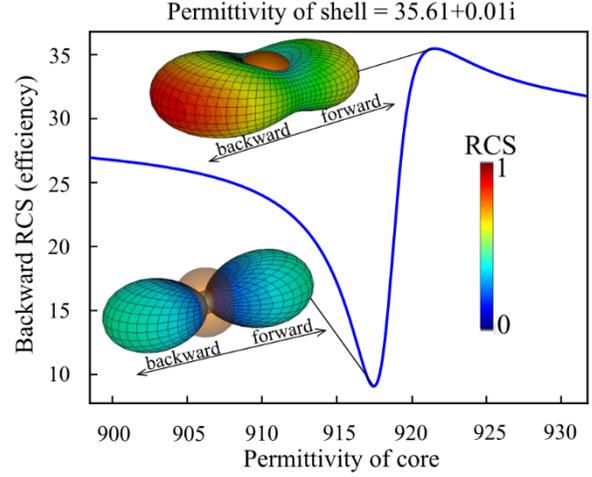

FIG. 3. Backward radiation cross section spectra depending on refractive index of core sphere. The refractive index of shell is constant. (Insert) Radiation pattern of core-shell structure in the points of maximal and minimal values of backward RCS.

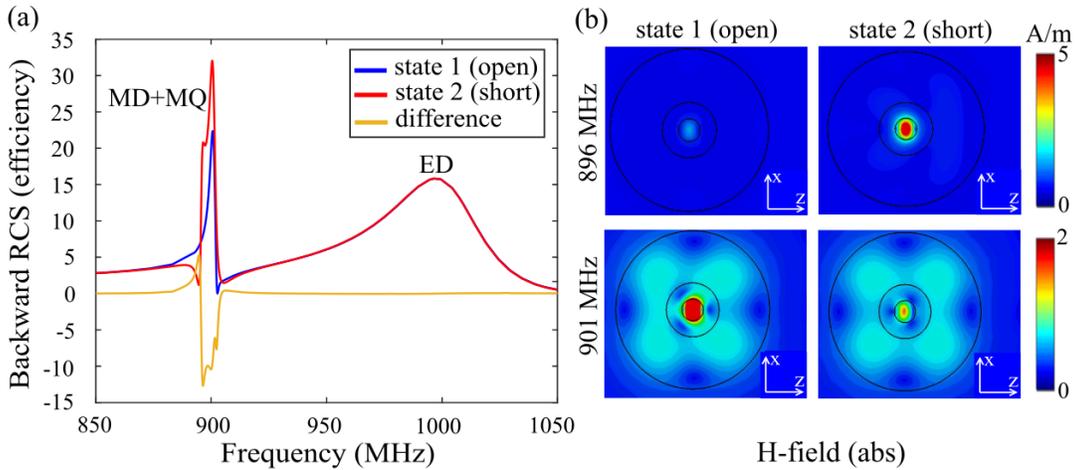

FIG. 4. (a) Backward scattering cross section spectra for an open circuit (blue curve), short circuit (red curve) and the difference between the states (orange curve). MD, MQ, and ED are the multipoles, contributing to the peaks. (b) Magnetic near field maps - absolute normalized values. Integrated circuit states and frequencies are indicated on the plot.

While in the previously considered theoretical model the resonance frequency is controlled by changing the permittivity of the inner sphere, this approach is not practical from the implementation standpoint. To shift magnetic dipolar resonance of the core we will introduce a split ring resonator (SRR), wrapped around it. An integrated circuit (IC) is plugged within the SRR's gap. Commercial RFID ICs typically have complex impedances in modulation states and may vary from vendor to vendor. To make the model general, we will assume the modulation between an open and short circuit. The layout of the device appears in Fig. 1 – SRR is inductively coupled to the inner sphere and, as a result, the backscattering strongly depends on the IC's state.

To analyze and optimize the reperformance of the structure we made a full-wave numerical simulation in CST Microwave Studio, frequency-domain solver. The same geometrical parameters, as in the theoretical model, have been used $r_{\text{core}} = 5.5$ mm, $\varepsilon_{\text{core}} = 898$ ($\tan \delta = 5 \cdot 10^{-4}$). SRR is made of copper wire (with losses), $r_{\text{ring}} = 6$ mm, the wire's diameter $d_{\text{wire}} = 1$ mm. The ring has two gaps, one of them is functionalized with the IC, the second one ($w_{gap} = 0.02$ mm) remains open to tune the SRR's impedance for the effective inductive coupling with the core sphere. The shell properties are: $r_{\text{shell}}^{\text{inner}} = 13.5$ mm, $r_{\text{shell}}^{\text{outer}} = 39.5$ mm, and $\varepsilon_{\text{shell}} = 35.7$ ($\tan \delta = 5 \cdot 10^{-4}$). The structure is excited by a plane wave (see Fig. 1 (a) for the configuration).

Numerically calculated backward scattering cross-section for the two IC's states are shown in Fig. 4 (a). The contribution of MD, MQ, and ED multipoles to the backscattering can be clearly identified. It can be seen that the combination of MD and MQ is most sensitive to the IC's state. The theoretical model also predicted that small perturbations will mainly affect the intersection of MQ and MD branches. The difference between the cross-sections [yellow curve in Fig. 4 (a)] reaches the absolute value of 12. Also, the backscattering does not vanish in either state, similarly to the theoretical case.

In order to link the far-field changes with the inductive near-field coupling at the inner volume of the structure, magnetic field amplitudes were calculated for both states of the IC and at two characteristic frequencies - $f = 901$ MHz and 896 MHz [see Fig 4 (b)]. At those frequencies, the backscattering difference reaches either maximal negative or positive values. The switching mechanism can be clearly seen now that the state of the IC controls the field localization in the core – the higher field localization is, the less energy is scattered to the far-field. This mechanism links the field localization strength with the sign of the backscattering difference.

Finally, the modulation efficiency degradation is assessed versus an introduction of all loss channels. Figure 5 demonstrates the far-field scattering diagrams (H-plane) for both IC's states in the case both ceramic elements and SRR are made of lossless materials (SRR is made of a perfect electric conductor). Dotted lines correspond to the lossy ceramics and copper as the SRR's material. The backscattering modulation efficiency is 21 and 12 geometrical cross-sections in lossless and lossy cases correspondingly. It means that no significant degradation occurs, and the concept can be implemented in real-life scenarios. It is also worth noting that at $f = 896$ MHz (the frequency for which Fig. 5 was plotted), open circuit state corresponds to almost complete suppression of MD and MQ and the scattering is governed by ED. The scattering diagrams almost perfectly resemble the dipolar scattering pattern, confirming the proposed analysis of the interaction dynamics.

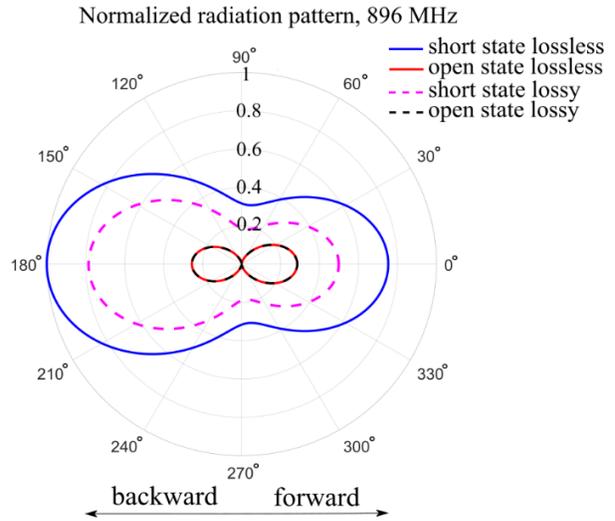

FIG. 5. Normalized radiation patterns in linear scale, polar plots (H-field planes).

In conclusion, the problem of backscattering modulation maximization was considered and applied to boost efficiencies of passive RFID communication channels. Multipole engineering, resembling generalized Kerker approach for scattering management, have been developed and applied on a core-shell geometry. It was shown that the backscattering suppression, typically targeted in Kerker-related problems, is a less efficient approach to achieve an enhanced backscattering modulation efficiency. On the other hand, controlling interference between higher-order multipoles is a preferable route. This concept was shown to provide a modulation efficiency of 41 geometrical cross-sections in the absence of losses and 26, when they were taken into account. The approach of multipole interference allows to design new devices for wireless communication systems with a broad range of applications across different frequency bands.

The research was supported by the Russian Science Foundation (Project 19-79-10232), ERC StG 'In Motion,' PAZY Foundation, and Israeli Ministry of Science and Technology (project "Integrated 2D & 3D Functional Printing of Batteries with Metamaterials and Antennas."